\pdfoutput 1
\documentclass[11pt,twoside,twocolumn,a4paper]{article}

\usepackage{cvww}
\usepackage{times}
\usepackage{epsfig}
\usepackage{graphicx}
\usepackage{amsmath}
\usepackage{amssymb}

\usepackage[table]{xcolor}
\usepackage{tabularx}
\makeatletter
\@namedef{ver@everyshi.sty}{}
\makeatother
\usepackage{pgfplots}
\pgfplotsset{compat=newest}
\usepgfplotslibrary{groupplots}
\usepgfplotslibrary{dateplot}

\usepackage{booktabs}
\usepackage{multirow}
\usepackage{makecell} % for more vertical space in cells
\usepackage{caption} 
\usepackage{subcaption}
\usepackage{mathptmx}
\usepackage{t1enc}

\cvwwfinalcopy % *** Uncomment this line for the final submission

\def\cvwwPaperID{12} % *** Enter the CVWW Paper ID here

\ifcvwwfinal\fi
\begin{document}

%%%%%%%%% TITLE
\title{CNN-CASS: CNN for Classification of Coronary Artery Stenosis Score in MPR Images}

\author{Mariia Dobko, Bohdan Petryshak, Oles Dobosevych\\
The Machine Learning Lab, Ukrainian Catholic University, Lviv, Ukraine\\
{\tt\small \{dobko\_m, petryshak, dobosevych\}@ucu.edu.ua}}

\maketitle
\ifcvwwfinal\thispagestyle{fancy}\fi

%%%%%%%%% ABSTRACT
\begin{abstract}
   To decrease patient waiting time for diagnosis of the Coronary Artery Disease, automatic methods are applied to identify its  severity using Coronary Computed Tomography Angiography scans or extracted Multiplanar Reconstruction (MPR) images,  giving doctors  a  second-opinion  on  the  priority  of  each case.  The  main  disadvantage  of  previous  studies  is the lack of large set of data that could guarantee their reliability. Another limitation is the usage of handcrafted features requiring manual preprocessing, such as centerline extraction. We overcome both limitations by applying a different automated approach based on ShuffleNet V2 network architecture and testing it on the proposed collected dataset of MPR images,  which  is  bigger  than  any  other used in this field before. We also omit centerline extraction step and  train and test our model using whole curved MPR images of 708 and 105 patients, respectively. The model predicts one of three classes: `no stenosis' for normal, `non-significant' --- 1-50\% of stenosis detected, `significant' --- more than 50\% of stenosis. We demonstrate model's interpretability through visualization of the most important features selected by the network. For stenosis score classification, the method shows improved performance comparing to previous works, achieving 80\% accuracy on the patient level. Our code\footnote{\url{https://github.com/ucuapps/CoronaryArteryStenosisScoreClassification/}} is publicly available. 
   
\end{abstract}

%%%%%%%%% BODY TEXT

\section{Introduction}

\begin{figure}[ht!]
\centering
{
\includegraphics[width=1\columnwidth]{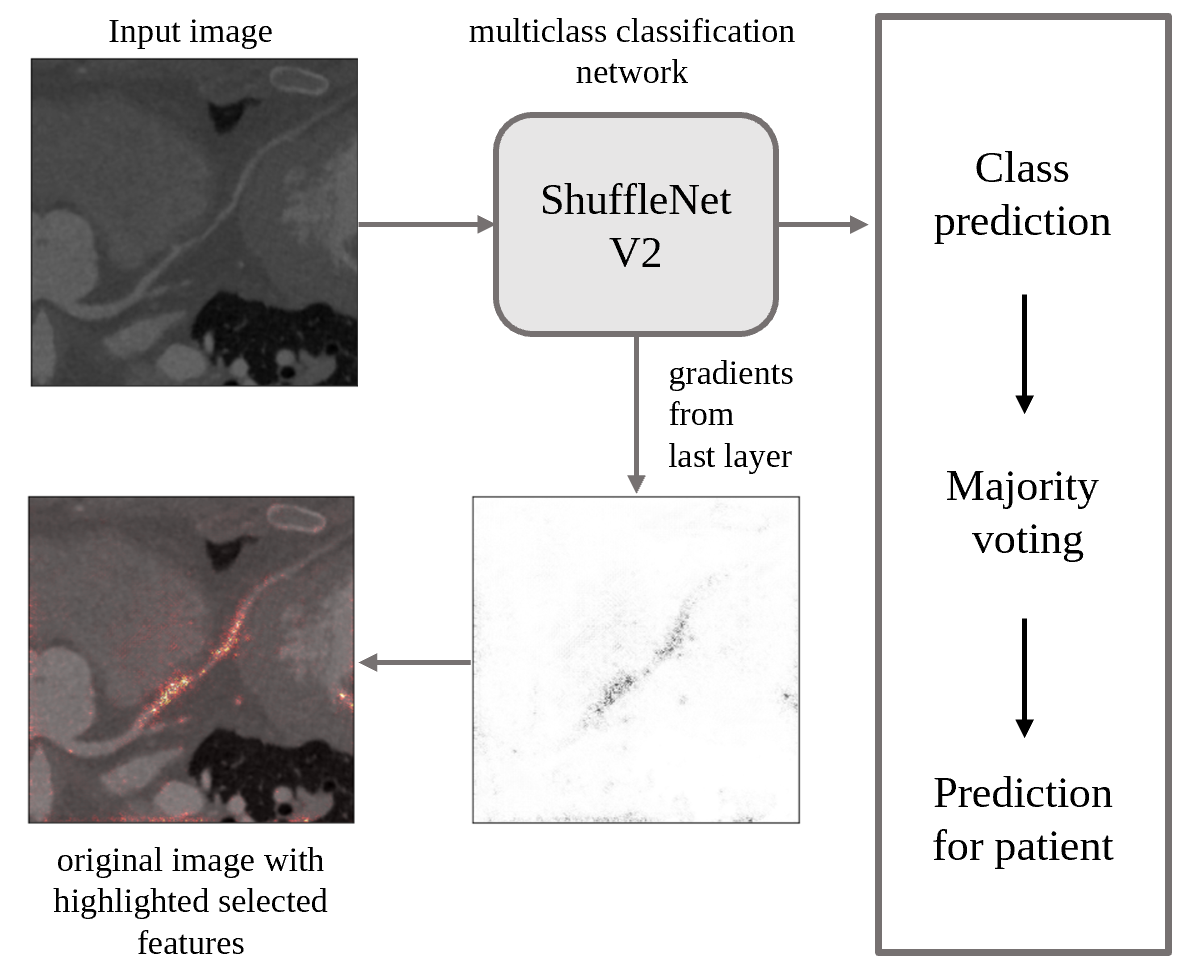}
}
\vspace{-1em}
\caption{
The pipeline of stenosis classification on an MPR image. For each 2D image, ShuffleNet V2  predicts probabilities of stenosis score. Activation regions of the last layer in the model are shown and overimposed on the input image.
\vspace{-1em}}
\label{fig:first}
\end{figure}

According to the American Heart Association report \cite{Benjamin2019}, approximately 17.6 million deaths were attributed to Cardiovascular Diseases (CVD) globally in 2016, making it the leading cause of death in the world. By 2030 it is estimated that CVD will be responsible for over 23.6 million deaths \cite{Benjamin2019}, and thus the ability to get early diagnosis becomes crucial. Cardiovascular diseases affect the heart or blood vessels. They include Coronary Heart Disease, or Coronary Artery Disease (CAD), which occurs when plaque (a combination of cholesterol, calcium, fat and other substances) builds up in the arteries and clogs them. This narrowing of the arteries, called stenosis, interferes in the healhy blood flow by hindering oxygen-rich blood cells transportation to the heart.

Coronary Computed Tomography Angiography (CCTA) is one of the components used by radiologists in diagnostics of the coronary artery disease. In the recent studies, CCTA was proven to be clinically effective in combination with functional testing (SCOT-HEART \cite{SCOT2015}), or even as an alternative to it (PROMISE \cite{Douglas2015}). Coronary CTA helps doctors in evaluation of the degree of artery stenosis. For patients in the risk group, it is vital to get the diagnosis in a short period of time, however, nowadays it takes up to two weeks to receive analysis results after the scanning procedure. In order to decrease the waiting time, the previous works \cite{kiricsli2013standardized}, \cite{zreik2018recurrent} applied various semi-automated algorithms to identify the severity of disease in medical images, giving doctors a second-opinion on the priority of each case. Such automated computer-aided systems are capable to increase the access to diagnostics and eventually reduce mortality by faster recognition of critical cases. 

Due to recent advances in applications of machine learning to medical domain, it is now possible to use neural networks for assessment of the severity of coronary artery stenosis. Among the main disadvantages of previous approaches are the lack of large set of data that could guarantee their reliability and usage of handcrafted features during the preprocessing steps. Our contribution to this field lies in application of a different approach and testing it on the created dataset, which is bigger than any other used in this field before. We also propose a fully automated method to classify stenosis score, that utilizes whole curved Multiplanar Reconstruction (MPR) images without manual preprocessing or centerline extraction, see Figure \ref{fig:first}.

\section{Related Works}

The problem of stenosis score classification on CCTA images of coronary arteries is insufficiently studied. The major difficulty is the absence of publicly available structured and professionally labeled sets of data.  Another one is domain specificity, which requires certain expertise in analysing medical data. 

Datasets of coronary arteries are usually formed by CCTA scans, from which MPR images can be extracted and transformed to either straight or curved representation of arteries. In the related paper [3], CCTA scans of only 163 patients were collected, and the proposed network was trained and tested using images of 98 and 65 patients, respectively. The authors first straightened MPR images by applying the centerline extraction technique \cite{Wolterink2019} and then used the transformed data to simplify classification of stenosis level. The centerline extraction step used in this approach requires manual placement of a single seed point in the artery of interest, so that the method is not fully automated. 3D convolutional neural network was utilized to extract features which are used by recurrent neural network for classification. While the achieved accuracy shows good performance and feasibility of deep learning methods for stenosis score classification, the reliability of obtained results can not be justified on such small data sample. Another drawback is its poor stability: as the authors admit, even small errors in centerline extraction may essentially increase the overall error. 

Centerline extraction is a common preprocessing method \cite{Metz2009} although it often requires manual assistance. It is used in the previous study \cite{zreik2018recurrent}, where user interaction is needed to localize the artery by annotating the start and end points of the vessel. As the authors described, the start point was placed in the coronary ostium of the corresponding arterial tree and the end point was placed at the most distal point inside the vessel. Some parameters were chosen manually, for example, contrast filled (foreground) regions are defined by empirically determining a lower and upper bound values of intensity. Other handcrafted features include the mean and dispersion values of the artery radius, as well as the mean and dispersion of a rotation angle corresponding to a typical location of the artery. 

In order to avoid errors caused by centerline extraction, in our approach we use curved MPR images instead. These MPRs are generated from CCTA with the help of radiologist assistant during a general pipeline of the coronary artery diagnosis. Thus, our method does not require any handcrafted features, but utilizes the whole MPR image, where artery is curved.

\begin{figure*}
\begin{center}
\includegraphics[width=\textwidth]{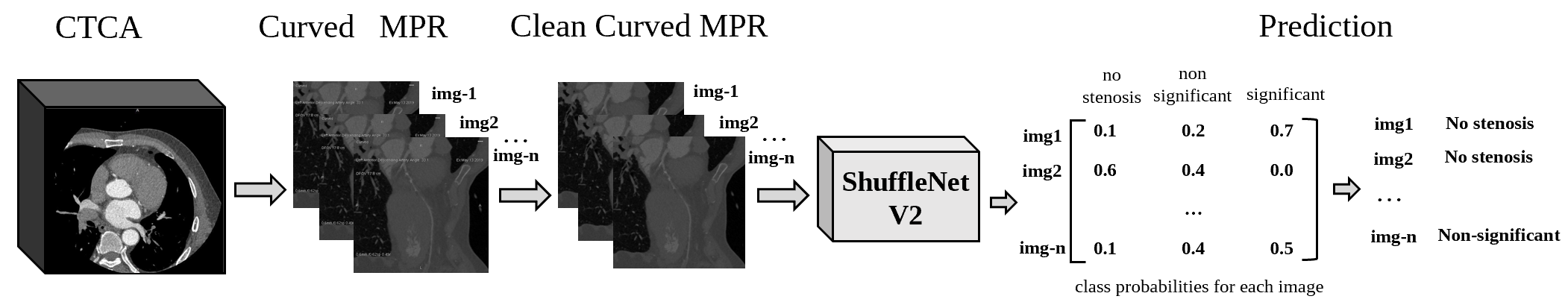}
\end{center}
   \caption{\textbf{Prediction process.} First, the whole CCTA is converted into sets of MPR images for each artery branch (completed by a radiologist assistant). Then, our method automatically cleans the image from the text and meta information, and feeds the obtained preprocessed images to the ShuffleNet V2. As an output, the probability predictions of each class are produced. }
\label{fig:predictionprocess}
\end{figure*}
\section{Data}

For training our algorithms, we used curved MPR images of the coronary artery with stenosis levels annotated by professional radiologists from a well-renowned Future Medical Imaging Group (FMIG) in Australia. Our current dataset consists of 160,000 MPR images which were extracted from CCTA scans of 828 unique patients; see data statistics in Table~\ref{tab:datastats}.

\begin{table}
   \centering
  \resizebox{0.8\linewidth}{!}{
   \begin{tabular}{ l l l c}
  \toprule

   Arteries & Arteries\# &Sections & Sections \\
   
  \toprule

   \textbf{LAD}  & 824 & \textbf{LAD} & 822 \\
   & & \textbf{D-1} & 729\\
   & & \textbf{D-2} & 356\\
   & & \textbf{D-3} & 68\\
   \midrule
   \textbf{LCX} & 722 & \textbf{LCX} & 639 \\
   & & \textbf{PLV-LCX} & 15\\
   & & \textbf{PDA-LCX} & 17\\
   \midrule
   
   \textbf{RCA} & 721 & \textbf{RCA} & 91 \\
   & & \textbf{OM} & 6\\
   & & \textbf{OM-1} & 81\\
   & & \textbf{OM-2} & 281\\
   & & \textbf{OM-3} & 75\\
   & & \textbf{PLV-RCA} & 609\\
   & & \textbf{PDA-RCA} & 71\\
   \bottomrule
   
   \end{tabular}
   }
   \vspace{1em}
   \caption{\textbf{Collected dataset statistics.} Number of cases containing certain arteries and branches for all 828 patients.}
   \vspace{-1em}
   \label{tab:datastats}
\end{table}
\subsection{MPR generation process} \label{mpr_generation_process}
The CCTA stack of images representing the patient's heart, is produced by the CT scanner. The raw CT scans require thorough and long analysis as the coronary artery is represented in each slice as a small circular region, similar to a dot. To increase the interpretability of the data, the clinicians use the MPR technique \cite{mpr_paper, mpr_second_article}.  MPR is the process of using the data from axial CT images to create a more anatomical representation of the coronary artery by tracking the whole specific artery branch along the CT volume and generating its two-dimensional image. 
\par
Each branch is represented by 50 MPR images, where one image corresponds to the specific view angle (180 degrees in total). The reason for that is that plaque might be located anywhere along the vessel, and be invisible only from some view angles. 

\subsection{Labels extraction from medical reports}
\label{labels}
The condition of coronary artery of each patient is described by the report. It contains the meta information about the person (age, gender, heart rate, etc.), characterization of stenosis score, type of the plaque and calcium score to all artery sections and branches. The raw reports are not suitable for training classification machine learning algorithms as they do not have any specific category attached to the particular image or at least to a stack of images. 
\par
We created the parsing pipeline, which takes the report of the patient as an input and extracts all information relevant to our task. The parsed data include the description of all important artery sections and branches with corresponding stenosis categories. The latter are grouped according to the standard defined by the Society of Cardiovascular Computed Tomography (SCCT) and Coronary Artery Disease - Reporting and Data System (CAD-RADS) \cite{SCCT_RADS}: 0\% - Normal, 1-24\%- Minimal stenosis or plaque with no stenosis, 25-49\% - Mild stenosis, 50-69\% - Moderate stenosis, 70-99\% - Severe stenosis, 100\% - Total Occlusion. In the reports, three main artery sections are presented: LAD (Left Anterior Descending Artery) with D-1, D-2, D-3 branches; RCA (Right Coronary Artery) with PDA-RCA, PLV-RCA branches; LCx (Left Circumflex Artery) with OM-1, OM-2, OM-3, LCx-PDA, LCx-PLV branches.

\subsection{Data labeling process} Specific recommendation for further patient treatment depends largely on the identified level of stenosis \cite{SCCT_RADS}. No further cardiac investigation is required unless moderate (50-69\%) or higher stage was reported. Preventive therapy and risk factor modification is suggested for minimal or mild stenosis. Due to these specific regulations, we assign one of the three classes for each MPR image: `no stenosis' for normal cases, `non-significant' --- 1-50\% of stenosis detected, `significant' --- critical cases where more than 50\% of stenosis is present and instant doctor's attention is required.

\subsection{Challenges in the dataset}
\begin{figure}[ht!]
\begin{center}
\includegraphics[width=1.0\linewidth]{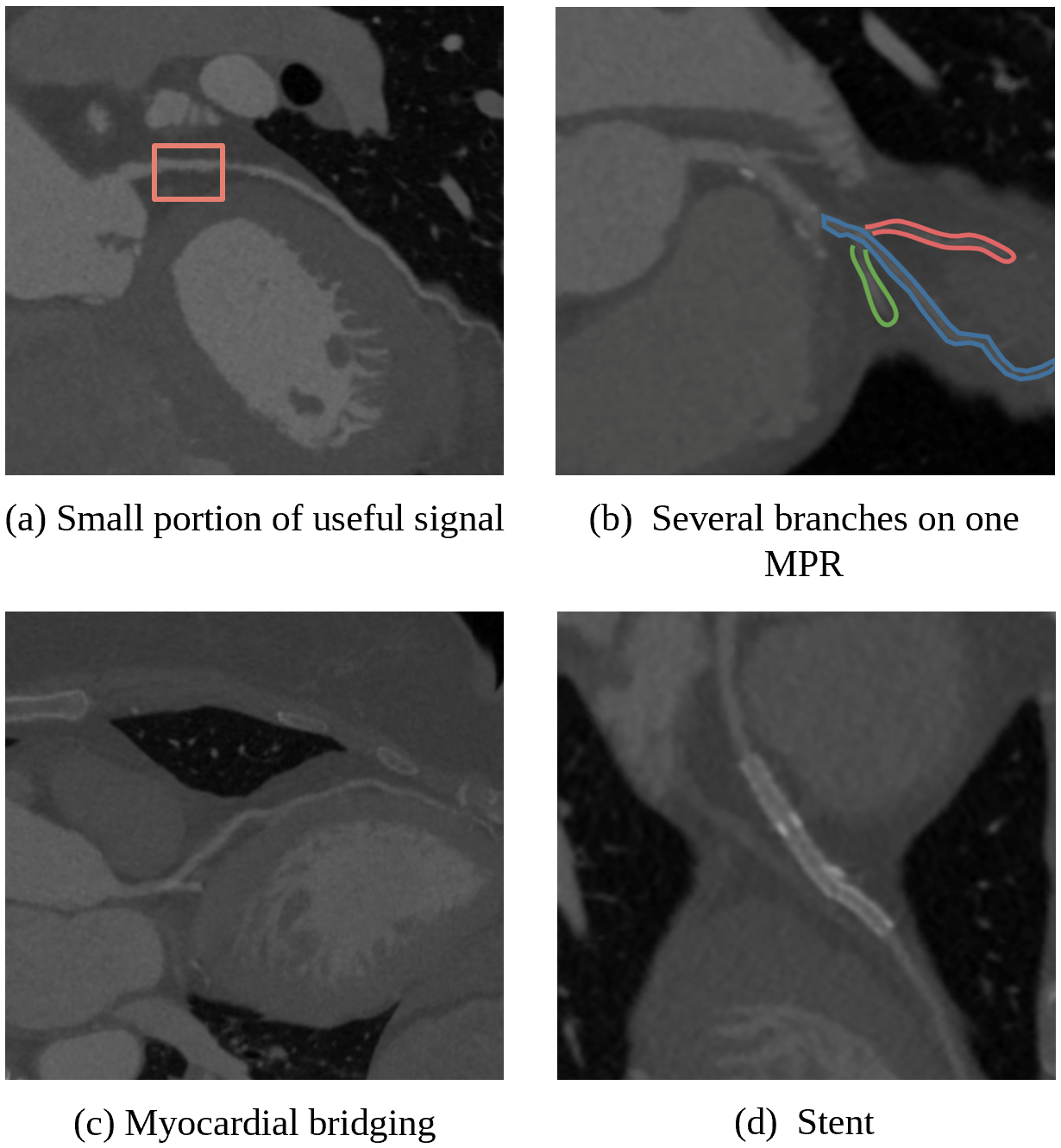}
\end{center}
\vspace{-1em}
   \caption{\textbf{Examples of difficult images.} Examples of the hard cases which are present in collected dataset. \textbf{(a)} The amount of pixels responsible for region of stenosis is many times smaller than the entire sample. \textbf{(b)} While the label for one MPR image corresponds to only one artery segment, several of them might be present on the image. \textbf{(c)} Physically natural narrowing can be visually similar to stenosis called myocardial bridging. \textbf{(d)} Example of artery with inserted stent, which can be mistakenly classified as stenosis with plaques.}
\label{fig:challenges}
\end{figure}

% mislabaled, ambiguous 
After all described preprocessing steps, we obtained a structured labeled dataset, but it is still incomplete and contains noise. One of the main issues is the appearance of several branches on the same image, see example in Figure \ref{fig:challenges} (b). This increases the risk of mislabeled data as each set of images representing one branch has just one label, which is then assigned to every single image in the set. For example, for the healthy LAD branch all the 50 images retrieved from different viewpoint angles are labeled as being normal. However, if on some of those images a neighbouring vessel (e.g., LCx)  with over 50\% score of stenosis is partly present, then the model may detect stenosis and misclassify a healthy LAD. 

Another issue is an inconsistency between medical reports written by doctors and the labeling system prescribed by CAD-RADS. The problem arises when the annotation provided by a radiologist is on the borderline between two classes from the CAD-RADS system. For example, the doctor might mark a specific branch by a "50\%"  of the stenosis. While it is satisfactory in medical terms, it becomes a challenge for us to choose which group this annotation belongs to - whether it should be considered as a significant or non-significant case.

Also we pre-process MPR images before inference. Each MPR image contains text information (meta-information about the picture), which has the highest intensity on the image. We remove it by assigning the average intensity of the neighbor pixels. Other examples of data challenges and difficulties are represented in Figure \ref{fig:challenges}.

\begin{figure*}
\begin{center}
\includegraphics[width=\textwidth]{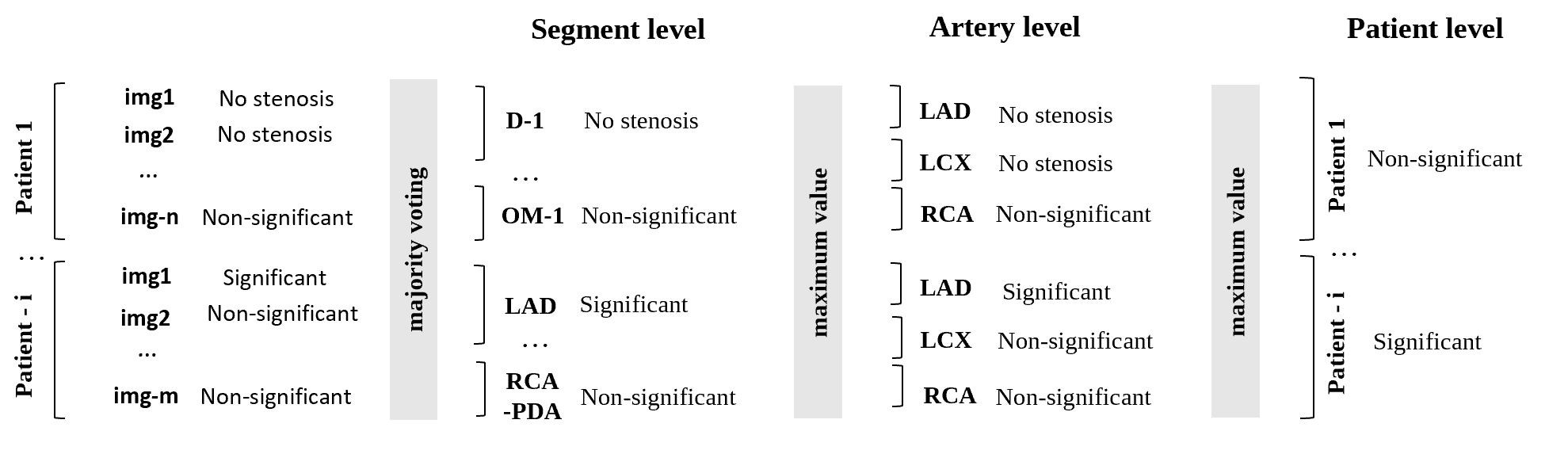}
\end{center}
  \caption{\textbf{Evaluation process.} The performance of proposed method is evaluated as suggested in \cite{vanRosendael2019} on segment-, artery- and patient-levels using $F_1$ score and accuracy. For each image, we predict one of the three classes, and then assign the class for the segment (branch) by applying the  majority rule to all 50 images corresponding to it. The maximal (most critical) class of all of the branches from one artery is assigned to this artery. The final prediction is calculated for each patient by choosing the maximal class out of all patient's arteries.}
\label{fig:evaluationprocess}
\end{figure*}

\section{Experiments}

\subsection{General pipeline}

We took 708 unique patients for training, 15 for validation and 105 for testing. We include the MPR images of every coronary artery and its branches described in Subsection~\ref{labels} in training and testing phases.
\par

We fed the MPR images into an optimized network architecture ShuffleNet V2. Each branch of the artery in most of the cases is represented by 50 MPR images (see Subsection~\ref{mpr_generation_process}). Thus for one branch, we get 50 predictions describing its stenosis score. Then using the majority rule, we assign the final stenosis score for the branch. The prediction pipeline is illustrated in~\autoref{fig:predictionprocess}, and it is followed by evaluation technique, shown in~\autoref{fig:evaluationprocess}.

%  Model blocks & architectures
\subsection{Methodology}

The technique which is widely used for optimizing the neural network architectures is 1x1 convolution. The authors of ShuffleNet \cite{Zhang2018} approached it and managed to reduce the time for this operation. The main idea behind ShuffleNet was to use separable depthwise convolutions \cite{xception}, grouped convolutions on 1x1 convolution layers - pointwise group, followed by channel shuffle operation. In this paper, we use an improved version of this architecture - ShuffleNet V2 \cite{Ma2018}, which is not only faster, but also more accurate.

We use ShuffleNet V2, pretrained on ImageNet \cite{imagenet}, to extract features from curved MPR images and classify them into three classes: `no stenosis', `non-significant stenosis', `significant stenosis'.

\begin{figure}[b]
\begin{center}
% \vspace{-2em}
\includegraphics[height=20em, keepaspectratio]{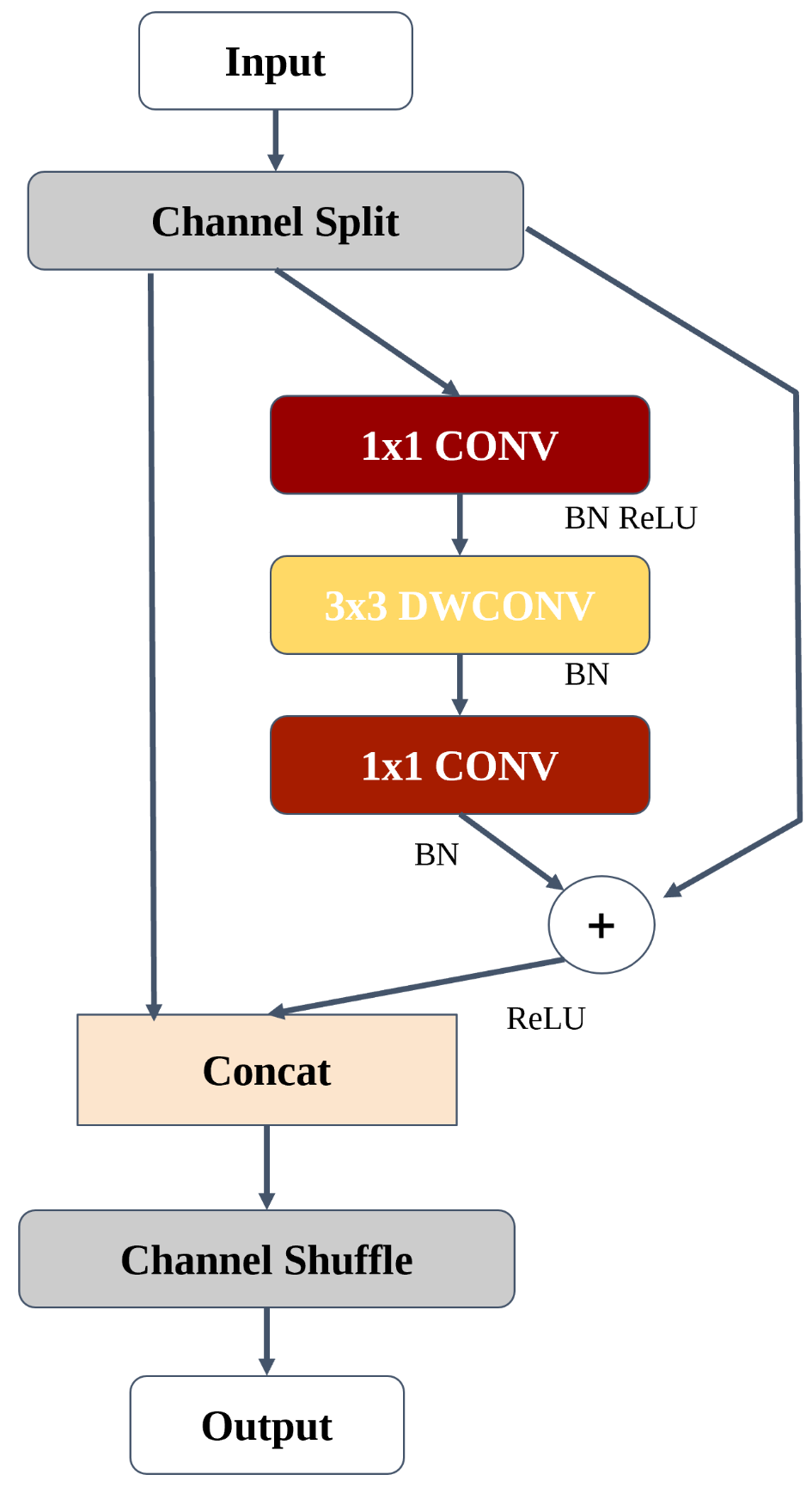}
\end{center}
   \caption{Structure of the basic building block of ShuffleNet V2 with residual \cite{Ma2018}. CONV: convolution layer. DWCONV: depthwise convolution. BN: batch normalization. Channel Shuffle: crucial operation for ShuffleNet architectures.}
\label{fig:shufflenet}
\end{figure}

The structure of the basic building block of ShuffleNet V2  \cite{Ma2018} with residual is displayed in Figure \ref{fig:shufflenet}. There are several building blocks which are stacked to construct the network. 
The input of feature channels is split into two branches at the beginning. In each block one branch directly goes through the block and joins the next one. The other branch has three convolution layers with the same input and output channels. Only one from the three 1x1 convolutions is group-wise. The two branches are concatenated after convolutions. For spatial downsampling the block is modified by the removal of split operator.  The Channel Shuffle  improves accuracy by enabling information communication between different groups of channels. 
% The high efficiency in each building block of ShuffleNet V2 \cite{Ma2018} enables using more feature channels and larger network capacity. 

% Optimization and learning rate ...
We trained the model on our dataset using Adam optimizer \cite{adam_optimizer} with $10^{-4}$ learning rate. We chose the best value of learning rate for our model using LR Range Test \cite{Smith2017}. It is a method that implies running the model for a few iterations with initially very small learning rate and then increasing it linearly between low and high learning rate values after each epoch. This allows to estimate the minimum and maximum boundary learning rates. The gradient accumulation and batch normalization \cite{batch_norm, gradient_accumulation} were used to increase the batch size and provide a better direction towards a minimum of the loss function. 

%augmentation
To introduce robustness properties and desired invariance in our model, we employed standard data augmentation techniques. In the case of MPR images, we primarily need a scale, rotate, blur, brightness, and transpose invariance. 

\subsection{Evaluation metrics}
The CAD-RADS classification is applied on a per-patient basis and represents the highest grade of stenosis from the coronary tree \cite{vanRosendael2019}. Taking this into account, we evaluate our method performance on segment-, artery- and patient-levels, and define stenosis score according to the maximum value found at the current level. On the artery level, the highest grade of stenosis is selected out of all the grades on segments. On the patient level, the maximal stenosis stage is chosen among all the arteries. The accuracy and $F_1$ score for multiclass classification are computed on each level. 

We apply weighted averaging for $F_1$ score measure because we deal with multiclass labels. For each label metrics are calculated and then average weighted by the number of true instances for each class:
% \[ F1(class_{1})*W_1 + F1(class_{2})*W_2 + F1(class_{3})*W_3 \]

\[\sum_{i=1}^{3} F_1(class_{i})*W_{i}\]

where $i$ - iterator over the 3 classes, $F_1$ is $F_1-score$ and $W$ - weight for the current class .

\begin{table}[hb!]
  \centering
%   \ra{1}
   \resizebox{1\columnwidth}{!} {
      \begin{tabular}{ccccc}
    % \begin{tabular}{@{} rc{6ex}C{10ex}C{} @{} }
      \toprule
          Results & \multicolumn{2}{c}{Accuracy} & \multicolumn{2}{c}{$F_1$ score} \\
      \toprule
      & Our & RNN-based\cite{zreik2018recurrent} & Our & RNN-based\cite{zreik2018recurrent} \\
      Segment-level &\textbf{0.81} & 0.80 & \textbf{0.81} & 0.75  \\
      Artery-level & \textbf{0.81} & 0.76 & \textbf{0.82} & 0.77 \\ 
      Patient-level & \textbf{0.80} & 0.75 & \textbf{0.80} & 0.75  \\
      \bottomrule
    \end{tabular}
   }
  \caption{Accuracy and $F_1$ score on test sets: 105 patients (approximately 25,000 MPR images in total) in our test, 65 patients in Zreik et al.\cite{zreik2018recurrent}}
\label{tab:metrics}
\end{table}

\begin{figure*}
\begin{center}
\includegraphics[width=0.95\textwidth]{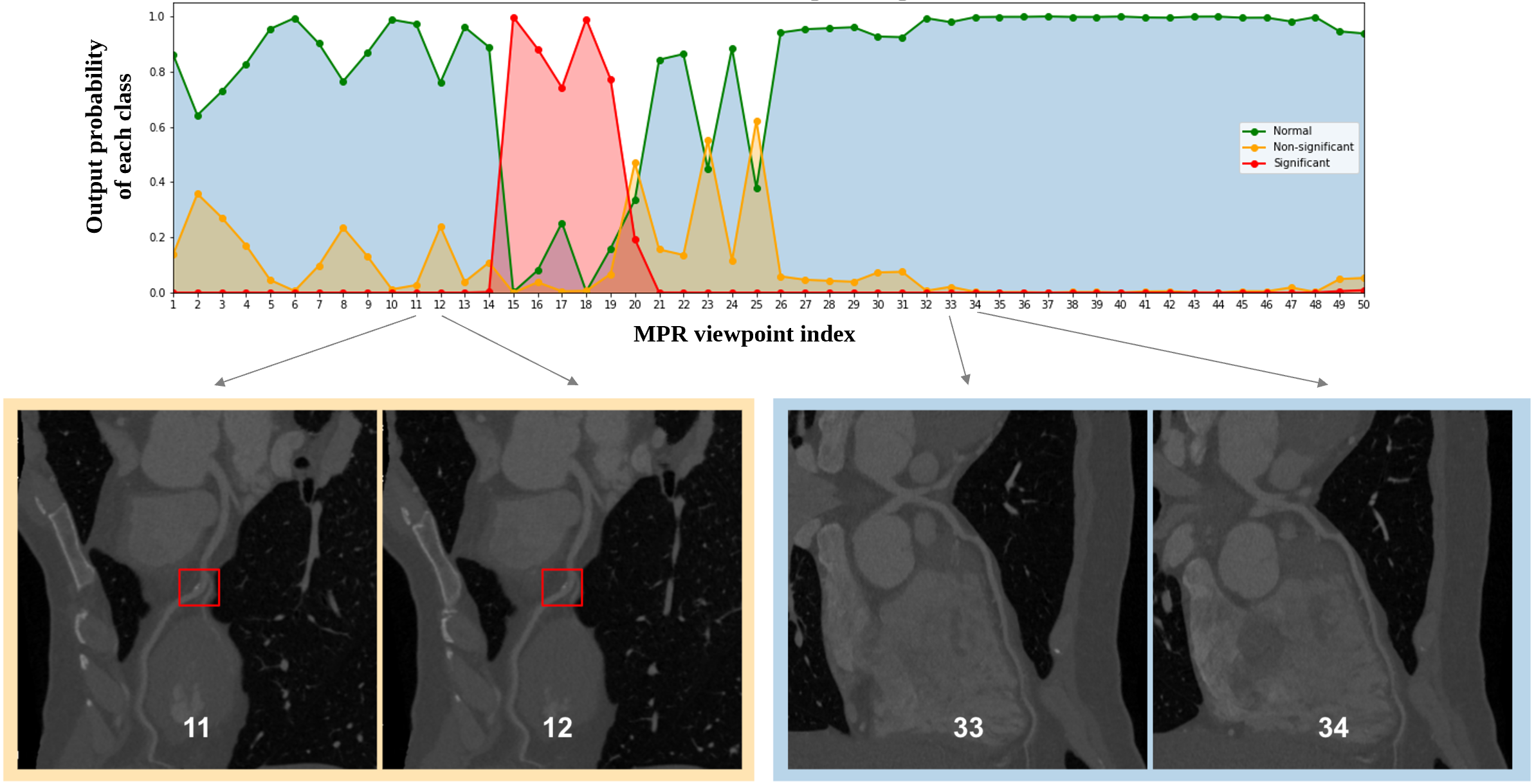}
\end{center}
  \caption{\textbf{Prediction for one segment.} The particular example has 25\% of stenosis in LAD artery. The labels for all 50 images representing this segment are the same - `non-significant'. While it is true for some of the MPR view points (see two images at the bottom left), from most of the angles stenosis is not seen either by a human eye, nor by a model (see two images at the bottom right). Thus, it is a direct illustration of weak labeling. It is important to treat every segment as a set of 50 images which are related. }
\label{fig:weaklabels}
\end{figure*}

The final results and comparison to previous study are reported in Table \ref{tab:metrics}. We also display confusion matrices in Table \ref{fig:cms} for every level separately in order to show results across the classes. This allows to observe the model's sensitivity (True Positive rate) - cases where the model correctly predicts the positive class, and specificity (True Negative rate) - cases where the model correctly predicts the negative class. For medical problems the False Positive error is always less dangerous than False Negative, these metrics are shown on confusion matrices. None of the patients with significant stenosis were classified as having no stenosis, and none of the healthy patients were put in the significant stenosis category. The model makes mistakes between `no stenosis' and `non-significant' classes, as well as between `non-significant' and `significant', which could be caused by the noise in data (see Figure \ref{fig:challenges}) and weak labels.   

\begin{table}[h]
{
\begin{subtable}[h]{1\columnwidth}%{0.45\textwidth}%{    % for making group where "\makegapedcells" is valid
\centering
\resizebox{1\columnwidth}{!}{ %
\begin{tabular}{cc|ccc}
\multicolumn{2}{c}{} & \multicolumn{3}{c}{Predicted} \\
%  & {\fontsize{11}{}\selectfont \textbf{Segment level}} & No stenosis & Non-Significant & Significant \\ 
& \textbf{Segment level} & No stenosis & Non-Significant & Significant \\ 
\cline{2-5}
\multirow{3}{*}{\rotatebox[origin=c]{90}{Actual}}
    & No stenosis   & \cellcolor{black!92}\color{white}0.92   & \cellcolor{black!7}0.07 & \cellcolor{black!1}0.01                 \\
    & Non-Significant    & \cellcolor{black!32}0.32    & \cellcolor{black!60}\color{white}0.6 & \cellcolor{black!8}0.08                \\ 
    & Significant    & \cellcolor{black!18}0.18    & \cellcolor{black!26}0.26  & \cellcolor{black!55}\color{white}0.55               \\ 
    \cline{2-5}
\end{tabular}
}
\end{subtable}
}
\vspace{1em}
{
\begin{subtable}[h]{1\columnwidth}%{0.45\textwidth}%{    % for making group where "\makegapedcells" is valid
\centering
\resizebox{1\columnwidth}{!}{ %
\begin{tabular}{cc|ccc}
\multicolumn{2}{c}{}
            &   \multicolumn{3}{c}{Predicted} \\
    & \textbf{Artery level}      &   No stenosis &   Non-Significant    & Significant          \\ 
    \cline{2-5}
\multirow{3}{*}{\rotatebox[origin=c]{90}{Actual}}
    & No stenosis   & \cellcolor{black!91}\color{white}0.91   & \cellcolor{black!8}0.08 & \cellcolor{black!1}0.01                 \\
    & Non-Significant    & \cellcolor{black!24}0.24    & \cellcolor{black!67}\color{white}0.67 & \cellcolor{black!9}0.09                \\ 
    & Significant    & \cellcolor{black!13}0.13    & \cellcolor{black!26}0.26  & \cellcolor{black!61}\color{white}0.61               \\ 

    \cline{2-5}

\end{tabular}
}
\end{subtable}
}
\vspace{1em}
{
\begin{subtable}[h]{1\columnwidth}%{0.45\textwidth}%{    % for making group where "\makegapedcells" is valid
\centering
\resizebox{1\columnwidth}{!}{ %
\begin{tabular}{cc|ccc}
\multicolumn{2}{c}{}
            &   \multicolumn{3}{c}{Predicted} \\
    &  \textbf{Patient level}    &   No stenosis &   Non-Significant    & Significant          \\ 
    \cline{2-5}
\multirow{3}{*}{\rotatebox[origin=c]{90}{Actual}}
    & No stenosis   & \cellcolor{black!81}\color{white}0.81   & \cellcolor{black!19}0.19 & \cellcolor{black!0}0.00                 \\
    & Non-Significant    & \cellcolor{black!7}0.07    & \cellcolor{black!80}\color{white}0.80 & \cellcolor{black!13}0.13                \\ 
    & Significant    & \cellcolor{black!0}0.00    & \cellcolor{black!21}0.21  & \cellcolor{black!79}\color{white}0.79               \\ 

    \cline{2-5}

\end{tabular}
}

\end{subtable}
}
  \caption{\textbf{Confusion matrices.} For each level: segment, artery and patient we calculate confusion matrix to see the number of False Positives, False Negatives and compare them.}
\vspace{-1em}
\label{fig:cms}
\end{table}

%  Results Visualization
\subsection{Results interpretability}
\begin{figure*}
\begin{center}
\includegraphics[width=0.9\textwidth]{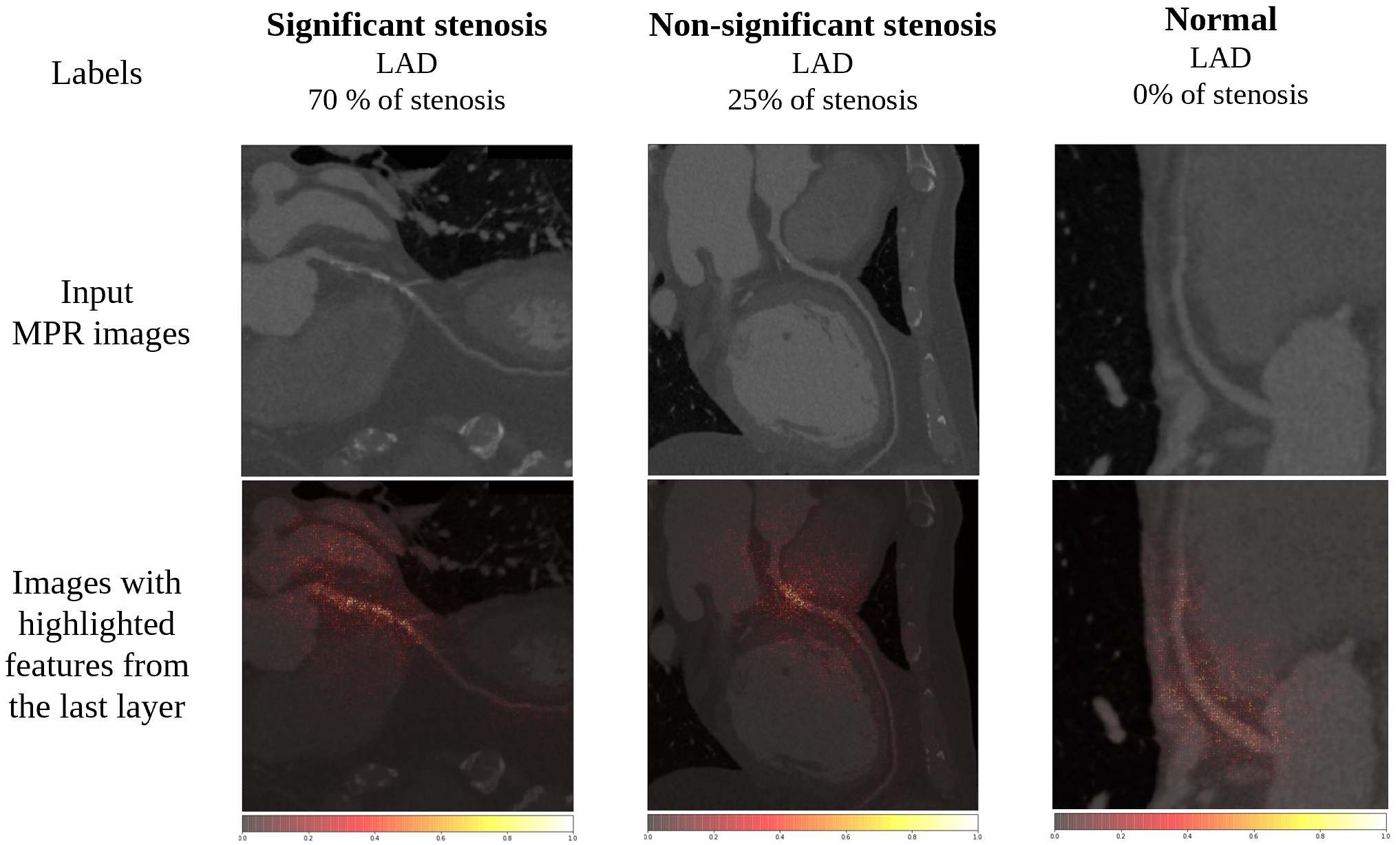}
\end{center}
  \caption{\textbf{Visualizations of the last layer's most important features for correctly predicted cases}. 
  These were created by using Integrated Gradients \cite{pmlr-v70-sundararajan17a}. Each class is represented for three examples of Left Anterior Descending artery: significant, non-significant, no stenosis. \textbf{Top}: The input two-dimensional MPR images of three different patients generated from CCTA scans.
  \textbf{Bottom}: The gradient visualization of most impactful features for the model. The brighter the pixels are, the more importance they have in prediction.}
\label{fig:visualizebest}
\end{figure*}

In order to achieve model interpretability we use Captum \cite{captum2019github} library containing implemented methods that identify which training features are important for the model. We visualized the features from the last layer of our model to understand which image regions have the largest impact on the model. The attribution of the network prediction to its input features was performed by applying axiomatic attribution method --  Integrated Gradients \cite{pmlr-v70-sundararajan17a} implemented in Captum \cite{captum2019github}. In this approach, the integral of the gradients of the output prediction for the specified class is computed with respect to the input image pixels. We observe that for all types of arteries and all levels of stenosis the model pays attention mostly to the artery zone, while the background with noise does not play a role in classification. Some examples of the features visualization using heatmaps are shown in Figure \ref{fig:visualizebest}. It is noticeable that for images which contain stenosis, model is more confident in the regions where plaques are located. This demonstrates model reliability, since plaques presence directly correlates with stenosis. 

\begin{figure}[ht!]
\begin{center}
\includegraphics[width=1.0\linewidth]{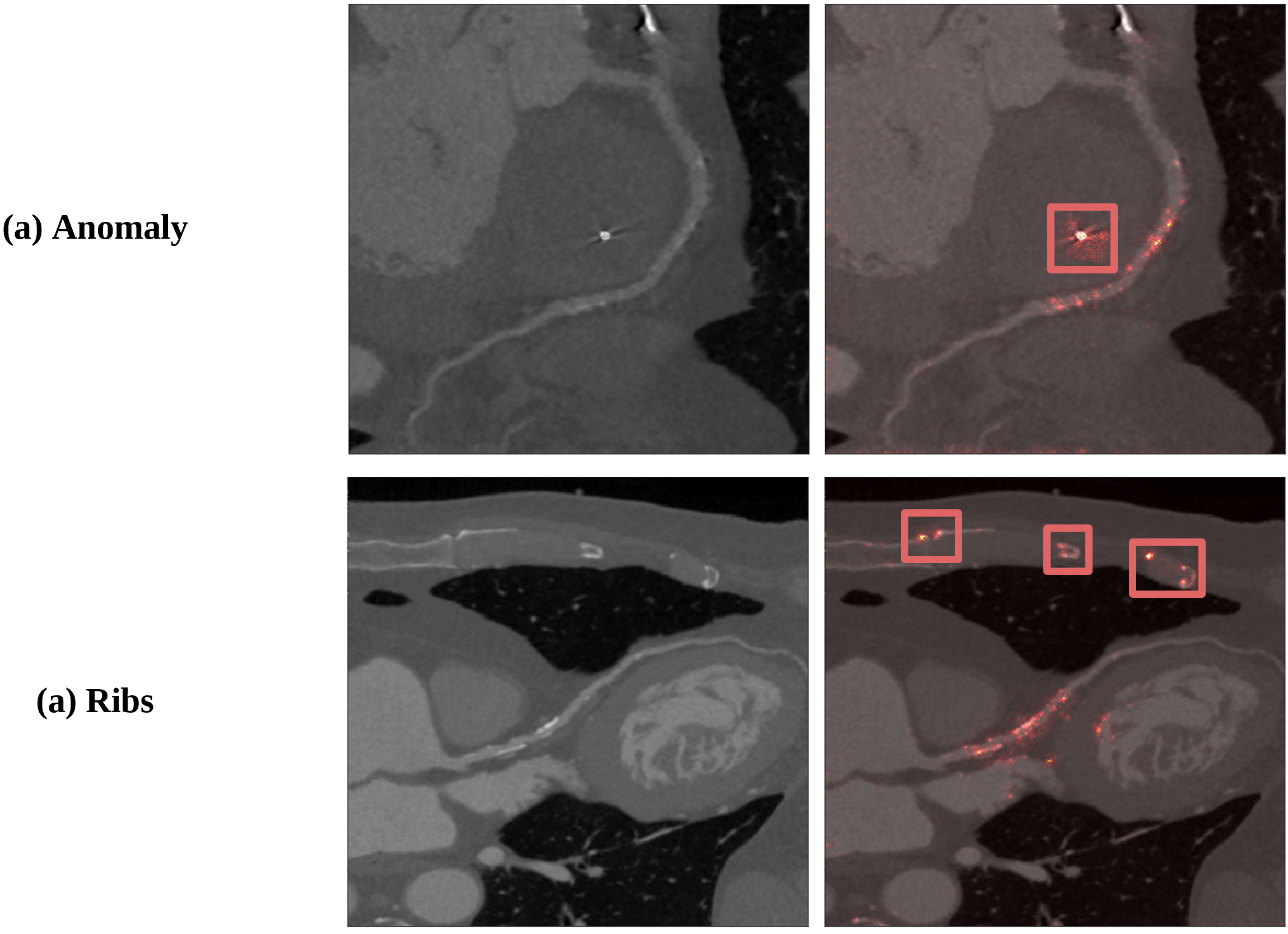}
\end{center}
\vspace{-1em}
  \caption{\textbf{Visualization of the model's confusion.}
  }
  Scans with the structures, which are similar to the calcified plaque. \textbf{(a)} The anomaly, with the high level intensity pixels. \textbf{(b)} Calcified ribs, which caused the model confusion.
\label{fig:viz_hard_cases}
\vspace{-1em}
\end{figure}

Although our model is capable of handling the background noise information and mainly takes into account only relevant areas, there are some corner cases. One of the most common types of plaque is calcified plaque. In computed tomography scans, it is represented as the pixels with a high level of intensity. There are structures and tissues like calcified ribs, sternum, costal cartilages, ventricle walls, etc., which are in the same range of radiodensity. The visualization showed that our models pay attention to these regions and associate them with the stenosis presence, which causes the lower specificity of our algorithms (see ~\autoref{fig:viz_hard_cases}).

\section{Conclusion}

We propose a simple automated framework, which is capable of detecting the stenosis score in curved MPR images. Our method shows improvements over the previous results \cite{zreik2018recurrent} reporting 80\% accuracy for the multiclass classification of stenosis level.

Our main contribution lies in creating new dataset of Cardiac CT scans of more than 800 patients, which is larger than any previous dataset, and suggesting a new approach for stenosis level classification. The proposed method omits centerline extraction and does not require any handcrafted features. Furthermore, we obtain explainable results and display features which impacted network's decisions.

The model interpretability through visualization of feature importance is very helpful in medical imaging as radiology specialists may use it to build trust in the model's predictions, refine classification of borderline cases, as well as gather observations for future testing.

\section{Discussion}

There are several ways to improve our approach. The curved MPR images, which were used as input in this work, contain not only artery, but also the background, where there is no useful information for determining the level of stenosis. We believe that the results of the proposed approach can be improved by using segmentation as an additional step during preprocessing and feeding the network with segmented images, where only the artery region is present.

Taking into account the problems shown on ~\autoref{fig:viz_hard_cases}, we might improve the performance of our models by adding the attention gate \cite{attention_gates} to the current network architecture. It will automatically learn the relevant areas for our task and suppress the unrelated target structures.

To obtain the final stenosis score for one branch, we take 50 predictions of our network for each corresponding MPR image representing the artery and assign the prevailing class. With this approach, we do not take into account the spatial relationship between MPR images. One possible improvement might be to apply the 3D CNNs to catch patterns across three spatial dimensions. One of the options to skip the step of MPR extraction is to create a new method, which would directly use 3D images of CCTA for stenosis score classification. 
\par

Due to the difficulty in collecting reliable labels for medical data,  unsupervised or weakly-supervised approaches should be considered. We believe that one of the possible ways of implementing such a solution is to train autoencoder exclusively on normal images with noise, then, to decide whether the particular MPR image represents the normal case based on its distance to the corresponding image generated by the model. 

The other area for research is the extraction of MPR images from CCTA. This task is handled semi-manually by a radiologist at the clinics, therefore, it is costly and takes long time. Our dataset already contains extracted MPR data. We think this process can be simplified by application of deep learning in building multiplanar reconstruction images based only on data from CCTA scans.

\section*{Acknowledgements}

This research was supported by Faculty of Applied Sciences at Ukrainian Catholic University. The authors thank Future Medical Imaging Group for providing the access to their data, Andrew Dobrotwir for constructive discussions and sharing medical expertise, Rostyslav Hryniv for helpful insights, and Jan Kybic for valuable feedback.

{\small
\bibliographystyle{ieee}
\bibliography{cvww_template}
}

\end{document}